\documentclass[aps,superscriptaddress,twocolumn]{revtex4}

\usepackage{graphicx}
\usepackage{amsmath}
\usepackage{amsfonts}
\usepackage{amssymb}

\providecommand{\unit}[1]{\,\mbox{#1}}

\begin{document}

\title{``Reservoir model'' for shallow modulation-doped digital magnetic quantum wells}

\author{Henrique J. P. Freire}
\email{freire@if.sc.usp.br}
\affiliation{Departamento de F\'{\i}sica e Inform\'{a}tica, Instituto de F\'{\i}sica de S\~{a}o Carlos, \\
Universidade de S\~{a}o Paulo, Caixa Postal 369, 13560-970 S\~{a}o
Carlos-SP, Brazil}

\author{J. Carlos Egues}
 \email{carlos.egues@unibas.ch}
 \affiliation{Department of Physics and Astronomy, University of Basel,
Klingelbergstrasse 82, CH-4056 Basel, Switzerland}

\date{July 13, 2002}

\begin{abstract}
Digital Magnetic Heterostructures (DMH) are semiconductor
structures with magnetic monolayers. Here we study electronic and
magneto-transport properties of \emph{shallow} modulation-doped
(ZnSe/ZnCdSe) DMHs with spin-5/2 Mn impurities. We compare the
``reservoir'' model, possibly relevant to shallow geometries, to
the usual ``constant-density'' model. Our results are obtained by
solving the Kohn-Sham equations within the Local Spin Density
Approximation (LSDA). In the presence of a magnetic field, we show
that both models exhibit characteristic behaviors for the
electronic structure, two-dimensional carrier density, Fermi level
and transport properties. Our results illustrate the relevance of
exchange and correlation effects in the study shallow
heterostructures of the group II-VI.
\end{abstract}

\maketitle

Digital Magnetic Heterostructures (DMHs) are state of the art
layered semiconductor structures in which (quasi-)two-dimensional
distributions of magnetic moments are restricted to equidistant
planes within a quantum well \cite{Crooker95,Awschalom99}.
Spin-dependent quantum effects are pronounced in these systems.
For example, in an external magnetic field the \textit{s-d}
exchange interaction \cite{Furdyna88} between itinerant electrons
and those of the magnetic impurities is responsible for a giant
Zeeman effect up to two orders of magnitude larger than the
ordinary one. These large energy splittings are readily observed
by magneto-photoluminescence \cite{Crooker95,egues-wilkins}. In
addition, the successful achievement of high-doping carrier
densities in \cite{Smorchkova96} has enabled magnetotransport and
magneto-photoluminescence measurements in these systems
\cite{Smorchkova97,PASPS1,PASPS2}.

In this work we investigate spin-dependent properties of
\emph{shallow} DMHs, Fig.~1(a). ``Shallow'' here implies that the
confining potential is weak enough, e.g. $\sim\,25\unit{meV}$ for
a $\sim\,10.5\unit{nm}$ wide well [Fig.~1(b)], so as to have only
a few confined subbands. For such geometries, the adjacent
\textit{n}-doped regions with densities
$\sim\,10^{17}\unit{cm}^{-3}$ provide enough carriers to fill up
all confined levels in the well, thus serving as electron
reservoirs. We describe these shallow DMHs using a ``reservoir
model'' with a constant Fermi level pinned to the chemical
potential of the \textit{n}-doped regions
\cite{Baraff81,Raymond94}. Such a model has also been applied to
GaAs/GaAlAs \cite{Burgt95,Takagaki97,Raymond99} and to ZnSe/ZnCdSe
\cite{Quinn00} systems. We calculate the magnetic-field dependent
subband structure of our shallow DMHs by using Density Functional
Theory within a Local Spin Density Approximation (DFT/LSDA). As
recently shown, exchange-correlation effects are important in
shallow II-VI wells \cite{Freire01}. We determine Landau-level fan
diagrams, two-dimensional electron densities, and in-plane
transverse resistivities.

\begin{figure}[ht!]
 \begin{center}
 \resizebox{7.0cm}{!}{ \includegraphics{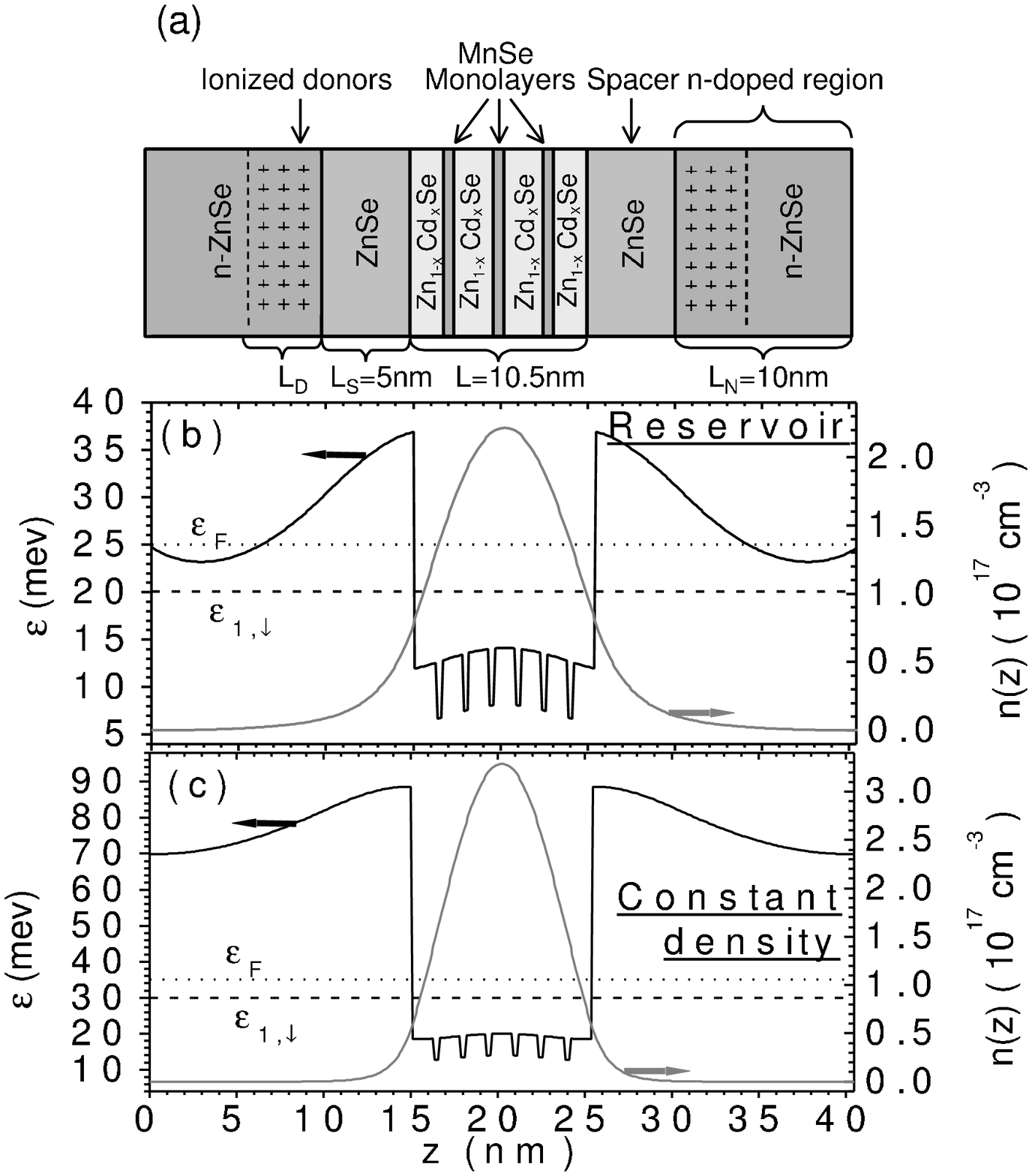}}
  \caption{(a) Layered structure of a digital magnetic quantum well
heterostructure. Lateral \textit{n}-doped regions (impurity
concentration of $1.2\times10^{17}\unit{cm}^{-3}$) provide
electrons that fill the confined energy levels. Self-consistent
potential profile for majority spin down electrons (at
$B=1\unit{T}$) are shown for the ``reservoir'' (b) and ``constant
density'' (c) regimes. Each case has only one confined subband
bellow the Fermi level. In (b) total electronic density $n(z)$
penetrates into the lateral barriers allowing exchange and
correlation to contribute to the potential profile far from the
well center.}
  \label{fig1}
 \end{center}
\end{figure}

\begin{figure*}[ht*]
 \begin{center}
  \includegraphics{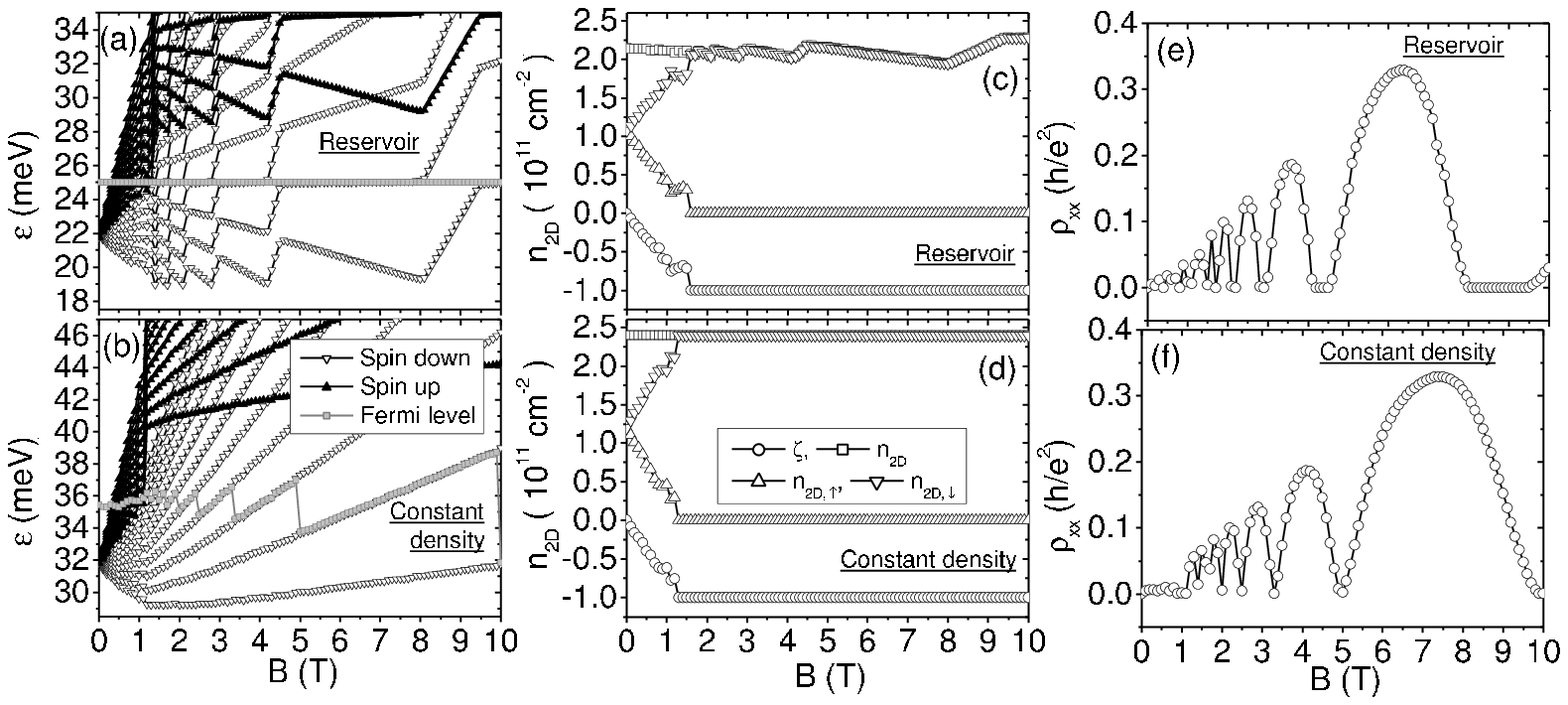}
  \caption{Magnetic field dependence of calculated Landau levels,
  two-dimensional densities, and transverse resistivities for both the
"reservoir" (upper panels) and the "constant density" (lower
panels) models. Within the reservoir model the Landau levels (a)
are pinned to the Fermi level $\varepsilon_F$ of the reservoirs,
thus originating sawtooth-profile LL fan diagrams. $\varepsilon_F$
(b) is pinned to the highest occupied LL in the constant-density
model and oscillates as the magnetic field is increased. The
\textit{s-d} enhanced LL spin splitting gives rise to the spin
polarization of the 2DEG at low B fields [(c) and (d)]. Note the
oscillatory behavior of the two-dimensional density within the
reservoir model (c). The transverse resisitivities $\rho_{xx}$
display Shubnikov-de Haas oscillations in both regimes (e) and
(f); however, plateaus $\rho_{xx}=0$ are seen only in (e) since we
do not account for localized states in this model. }
  \label{fig2}
 \end{center}
\end{figure*}

We compare our results within the ``reservoir model'' with those
obtained with the conventional ``constant-density'' model, in
which deep enough wells can confine \textit{all} available
carriers. In this case the \textit{n}-doped regions adjacent to
the well are fully depleted. We configure the system geometry for
each case by suitably choosing the height of the confining
barriers at the ZnSe/(Zn,Cd)Se interfaces: $70\unit{meV}$ and
$25\unit{meV}$ for the ``constant density'' and the ``reservoir''
models, respectively. The idea is to identify relevant contrasting
features in these models to eventually better describe real
shallow DMHs.


We calculate the magnetic-field dependent subband structure of our
system within Density Functional Theory (DFT). We solve the
Kohn-Sham (KS) equations in the Local Spin Density Approximation
(LSDA) using the parametrization of VWN \cite{Vosko80}. We compare
our LSDA results to those obtained within the Hartree
approximation to assess the role of exchange and correlation (XC)
effects in the properties of these novel systems. We consider the
system at zero temperature and perform the calculations in the
effective-mass approximation framework. We also make the usual
assumption that we can interpret the resulting KS eigenvalues as
subband energies of the system. Consequently, the total energy of
each Landau level (LL) is the sum of the subband energy
$\varepsilon_{i,\sigma_{z}}$ and the Landau cyclotron energy
$(n+1/2)\hbar\omega_{c}$ \nobreak
\begin{equation}
\varepsilon_{i,n,\sigma_{z}}(B) =\varepsilon_{i,\sigma_{z}}(B)
+\left( n+\frac{1}{2}\right) \hbar\omega_{c}+\frac{1} {2}g\mu_{\rm
B}\sigma_{z}B. \label{eq1}
\end{equation}
In (\ref{eq1}), the subband energy $\varepsilon_{i,\sigma_{z}}(B)$
corresponds to the longitudinal confined electronic motion and the
second term to the quantized transversal motion due to the
magnetic field. For completeness we also add the Zeeman energy,
although its contribution is very small when compared to the
\textit{s-d} exchange interaction.

Panels (b) and (c) of Fig.~1 show the resulting self-consistent
potential profiles for both ``reservoir'' (a) and
``constant-density'' (b) regimes. As the magnetic field increases,
the self-consistent potential and the corresponding electronic
structure change due to \textit{i}) magnetic-field dependent
contributions to the confining potential and \textit{ii}) the
rearrangement of electrons in the confined levels (magnetic-field
dependent Landau-level degeneracy $eB/h$).

Figures 2(a) and (b) show the Landau-level fan diagrams for both
regimes and the magnetic field dependence of the Fermi energy
$\varepsilon_{F}$. In the ``reservoir'' regime case the Fermi
level is constant at the chemical potential of the reservoir
(\textit{n}-doped regions) while in the ``constant density'' case
$\varepsilon_{F}$ oscillates as a function of $B$. In the first
case, all confined electronic levels are fully occupied because
\textit{n}-doped regions can always provide enough electrons to
the QW, thus being only partially depleted. For this reason the
LLs are pinned to the Fermi level (see plateaus at
$\varepsilon_{F}$). In the ``constant density'' case the
\textit{n}-doped regions are completely depleted but the amount of
electrons provided to the QW is not sufficient to fill all
electronic levels. Consequently, the Fermi energy is pinned to the
highest occupied LL, and oscillates as a function of the magnetic
field [Fig.~2(b)]. Note that in this case $\varepsilon_{F}$ is
pinned to the LL, not the contrary as in the ``reservoir'' regime.
This is why the curves in former case exhibit a sawtooth profile
not present in the latter. Note also that in the ``reservoir''
(``constant density'') case the two-dimensional density $n_{2D}$
changes (is constant) as a function of $B$ [Fig.~2(c) and (d)].

When compared to pure Hartree calculations (c.f. Ref.
\cite{Freire01}) we note that XC plays an important role in such
group II-VI heterostructures. Within LSDA calculations the LL spin
splitting is enhanced, the two-dimensional density is increased,
and the magnetic field at which the 2DEG becomes fully polarized
is decreased.

\bigskip

We also calculate the magnetic field dependence of the in-plane
resistivities (perpendicular to the magnetic field and growth
axis). We use Ando's model for the longitudinal conductivity
$\sigma_{xx}$ \cite{Ando74} and Drude's model for the transversal
component of the conductivity, $\sigma_{xy}=en_{2D}/B$. These
simple models enable us to illustrate general magnetotransport
features within these \textit{n}-doped magnetic structures and to
compare the two regimes for the density. In Fig.~2(e) and (f) we
plot our results for the in-plane longitudinal resistivity
($\mathbf{\rho}=\mathbf{\sigma}^{-1}$) which is a relevant
experimental quantity.

 Both ``reservoir'' and ``constant density''
regimes display the usual oscillations in $\rho_{xx}$
(Shubnikov-de Haas oscillations), but the former has regions of
zero resistivity [Fig.~2(e)].
In the ``reservoir'' regime the QW is always filled with the
maximum number of electrons possible, which is defined by both the
magnetic-field dependent density of states and the reservoir
chemical potential. Within this self-consistent equilibrium, there
are magnetic field ranges at which the center of the LLs are not
on the Fermi level [Fig.~2(a)], and consequently, the resistivity
is minimum [Fig.~2(e)].
In the ``constant density'' case, on the other hand, the Fermi
level is \textit{always} pinned to highest occupied LL and the
minimum resistivity dips [Fig.~2(f)] occur at magnetic fields that
$\varepsilon_{F}$ jumps between LLs [Fig.~2(b)]. At those magnetic
fields, $\varepsilon_{F}$ lies on the tail of a LL \cite{Tail} so
that there is only a small number of conducting electrons and,
consequently, a correspondingly small resistivity. In passing, we
note that in the standard model for the integer quantum hall
effect (IQHE) \cite{Klitzing80}, localized states in broadened LLs
are responsible for the Fermi-level pinning \cite{Stormer83}; the
phenomenological inclusion of these states (gaussian density of
states) gives rise to plateaus $\rho_{xx}=0$.

Finally, we should mention that we are currently investigating the
temperature dependence of the transverse magnetoresistivity
\cite{Knobel02}; a more detailed analysis, beyond the scope of
this communication, will be addressed in a future publication.
This additional study should allow us to contrast the two models
discussed here in terms of agreement with experimental data.

The authors thank Klaus Capelle for useful discussions. JCE
acknowledges support from the Swiss NSF, NCCR Nanoscience, DARPA,
and ARO. HJPF acknowledges financial support from FAPESP (Brazil).


\begin{thebibliography}{1}
\bibitem{Tail}
We actually simulate discrete Landau levels by using very narrow
($\sim\,0.004\unit{meV}$) gaussian-shaped profiles in the density
of states. Consequentely, $\rho_{xx}$ in Fig.~\ref{fig2}(e) and
(f) have a smooth oscilatory behavior.
\end{thebibliography}
\end{document}